\def\citen#1{\if@filesw \immediate\write \@auxout {\string\citation{#1}}\fi%
\@tempcntb\m@ne \let\@h@ld\relax \def\@citea{}%
\@for \@citeb:=#1\do {\@ifundefined {b@\@citeb}%
    {\@h@ld\@citea\@tempcntb\m@ne{\bf ?}%
    \@warning {Citation `\@citeb ' on page \thepage \space undefined}}%
    {\@tempcnta\@tempcntb \advance\@tempcnta\@ne
    \setbox\z@\hbox\bgroup\ifcat0\csname b@\@citeb \endcsname \relax
    \egroup \@tempcntb\number\csname b@\@citeb \endcsname \relax
    \else \egroup \@tempcntb\m@ne \fi \ifnum\@tempcnta=\@tempcntb
    \ifx\@h@ld\relax \edef \@h@ld{\@citea\csname b@\@citeb\endcsname}%
    \else \edef\@h@ld{\hbox{--}\penalty\@highpenalty
    \csname b@\@citeb\endcsname}\fi
    \else \@h@ld\@citea\csname b@\@citeb \endcsname \let\@h@ld\relax \fi}%
\def\@citea{,\penalty\@highpenalty\hskip.13em plus.13em minus.13em}}\@h@ld}
\def\@citex[#1]#2{\@cite{\citen{#2}}{#1}}%
\def\@cite#1#2{\leavevmode\unskip\ifnum\lastpenalty=\z@\penalty\@highpenalty\fi%
  \ [{\multiply\@highpenalty 3 #1%
  \if@tempswa,\penalty\@highpenalty\ #2\fi}]}   %
\makeatother 

\def\A             {\mbox{$\mathfrak A$}}
\def\AC            {\mbox{$\mathfrak C$}}
\def\ado           {\mbox{${\rm ad}^{}_{{\mfont U}_\omega}$}}
\def\adu           {\mbox{${\rm ad}^{}_{\mfont U}$}}
\def\aff           {affine Lie algebra}
\newcommand\al[1]  {{\alpha^{(#1)}}}
\newcommand\ala[2] {a_{#1}^{#2}}
\def\alam          {{a(\lambda)}}
\def\aLam          {{a(\Lambda)}}
\def\alamc         {{a(\lambda+c\delta)}}
\def\alamo         {{a(\omt\lambda)}}
\newcommand\alb[1] {{\bar\alpha^{(#1)}}}
\def\alg           {algebra}
\newcommand\alv[1] {{\alpha^{(#1)\Vee}}}
\def\aum           {\mbox{\mfont M}}
\def\auM           {{\mfont M}}
\newcommand\Aum[2] {\aum^{#1}_{\;#2}}
\def\aumm          {\aum^{-1}}

\def\aumt          {\aum^{\rm t}}

\newcommand\av[1]  {a^\Vee_{#1}}
\def\be            {\begin{equation}}
\def\bea           {\begin{eqnarray}}
\def\bearl         {\begin{array}{l}}
\def\bearll        {\begin{array}{ll}}
\def\bearlll       {\begin{array}{lll}}
\def\bfe           {{\bf1}}
\renewcommand\c[1] {\gamma_{#1}}
\def\Cdot          {\,{\cdot}\,}
\def\CDot          {{\cdot}}
\def\cdt           {\raisebox{.13em}{\tiny$\bullet$}}
\def\cft           {conformal field theory}
\def\Cft           {Conformal field theory}
\def\cfts          {conformal field theories}
\def\ci            {\cite{fusS3}}
\def\cii           {\cite{geni3}}
\def\Circ          {\,{\circ}\,}
\newcommand\cit[1] {\cite[#1]{fusS3}}
\newcommand\citq[3]{\cite[eq.\,(#1.#2#3)]{fusS3}}

\def\complex       {{\dl C}}

\def\coro          {\nsl{co}root}
\def\cwb           {Car\-tan\hy$\!$Weyl basis}
\def\daum          {{\mfont D}_{\mfont M}}
\newcommand\db[2]  {\bar\delta^{}_{#1,#2}}
\newcommand\dd[2]  {\delta^{}_{#1,#2}}
\def\df            {\,{:=}\,}

\def\dl            {\mathbb }

\def\dyd           {Dynkin diagram}
\def\ee            {\end{equation}}
\def\eE            {{\rm e}}
\def\eear          {\end{array}}
\def\Ei            {E^{\al i}_{}}
\def\Eipm          {E^{\pm\al i}_{}}
\def\Eib           {E^{\alb i}_0}
\def\Eibpm         {E^{\pm\alb i}_0}

\def\Eiopm         {E^{\pm\al 0}_{}}
\def\el            {\ell}
\def\Emi           {E^{-\al i}_{}}
\def\Emib          {E^{-\alb i}_0}

\def\Emob          {E^{\bar\theta}_{-1}}
\def\Emomdmi       {E^{-\al{\omdm i}}_{}}
\def\Eo            {E^{-\bar\theta}_1}
\def\Eopm          {E^{\mp\bar\theta}_{\pm1}}

\def\Eomdipm       {E^{\pm\al\omdi}_{}}
\def\Eomdmi        {E^{\al{\omdm i}}_{}}
\newcommand\eps[2] {e^{}_{#1,#2}}
\def\eq            {\,{=}\,}
\def\er            {{\bar\alpha}}
\newcommand\erf[1] {(\ref{#1})}
\newcommand\Erf[2] {(\ref{#1#2})}
\def\eri           {{\bar\alpha^{(i)}}}
\def\erI           {{\bar\alpha^{(i)}_{}}}

\def\erJ           {{\bar\alpha^{(j)}_{}}}
\def\es            {{\bar\beta}}
\def\ETA           {{\epsilon}}

\def\fline         {{~}\\[1 mm]\noindent ------------------\\[1 mm]}
\def\fock          {\mbox{$\cal F$}}
\newcommand\Frac[2]{\mbox{\large$\frac{#1}{#2}$}}

\def\futnot#1      { \ifnum\draftcontrol=1 \footnote{~{\sc internal footnote:}
                   #1}\addtocounter{footnote}{-1}\fi}
\def\futnote#1     {\footnote{~#1}\ }
\def\ihwm          {irreducible highest weight module}
\def\ihwr          {irreducible highest weight \rep}
\def\ii            {{\rm i}}
\def\imu           {\mu}
\def\infdim        {infinite-dimensional}
\def\iN            {\!\in\!}

\def\inu           {\nu}
\def\irmod         {irreducible module}

\def\izi           {{\displaystyle\oint\!\Frac{{\rm d}z}{2\pi\ii}\,}}
\def\g             {\mbox{$\liefont g$}}
\def\gb            {\mbox{$\bar{\liefont g}$}}
\newcommand\Gb[2]  {\bar G^{}_{#1,#2}}
\def\GB            {\bar G}
\def\gen           {\mbox{\mfont m}}
\def\gent          {\gen^{\rm t}}
\def\Gm            {G^{-1}}
\def\gm            {\mbox{$\liefont g_-^{}$}}
\def\gM            {{\liefont g}}
\def\go            {\mbox{$\liefont g_\circ^{}$}}
\def\gO            {{\liefont g_\circ}}
\def\god           {\mbox{$\liefont g_\circ^\star$}}
\def\goD           {{\liefont g_\circ^\star}}
\def\godb          {\mbox{$\bar{\liefont g}_\circ^\star$}}
\def\h             {\mbox{$\cal H$}}
\def\half          {\mbox{$\frac12\,$}}
\def\hgw           {highest \g-weight}
\def\hl            {\mbox{${\cal H}_\Lambda$}}
\def\hL            {{{\cal H}_\Lambda}}
\def\hlo           {\mbox{${\cal H}_{\omt\Lambda}$}}
\def\hLo           {{{\cal H}_{\omt\Lambda}}}
\def\hM            {{{\cal H}}}

\def\hw            {highest weight}
\def\hwm           {highest weight module}
\def\hwr           {highest weight representation}
\def\hwv           {highest weight vector}
\def\hy            {$\mbox{-\hspace{-.66 mm}-}$}

\def\kma           {Kac\hy Moo\-dy algebra}
\def\kv            {\mbox{$\el$}} 
\def\L             {\mbox{$\cal L$}}
\long\def\labl#1   {\label{#1}\ee \ifnum\draftcontrol=1
                   \mbox{ }\\[-3em]\query{#1}\\[1.5em] \fi}
\long\def\Labl#1#2 {\label{#1#2}\ee \ifnum\draftcontrol=1
                   \mbox{ }\\[-3em]\query{#1#2}\\[1.5em] \fi}
\long\def\labla#1  {\label{#1}\end{eqnarray} \ifnum\draftcontrol=1
                   \mbox{ }\\[-3em]\query{#1}\\[1.5em] \fi}

\def\Ldots         {,...\,,}
\def\Lg            {L_\gM}
\def\lhs           {left hand side}
\def\li            {\mbox{$\Lambda_{(i)}$}}
\def\lib           {\mbox{$\bar\Lambda_{(i)}$}}
\newcommand\Lib[1] {\mbox{$\bar\Lambda_{(#1)}$}}
\def\lie           {Lie algebra}

\def\liefont       {\mathfrak }

\def\llb           {\mbox{\large(}}

\def\lo            {\mbox{$\Lambda_{(0)}$}}
\def\LO            {\mbox{${\rm O}_L(r{+}1{,}1)$}}
\def\loo           {\mbox{$\Lambda_{(\omdo)}$}}
\def\loob          {\mbox{$\bar\Lambda_{(\omdo)}$}}
\def\LOp           {\mbox{${\rm O}^\uparrow_L(r{+}1{,}1)$}}

\def\lrb           {\mbox{\large)}}

\def\mer           {{m_\er}}
\def\mfont         {\rm }
\def\Mid           {\!\mid\!}
\def\munu          {{\imu,\inu}}
\def\Ne            {\,{\not=}\,}

\newcommand\normord[1] {\,\raisebox{.033em}{\large\bf:}#1
                   \raisebox{.033em}{\large\bf:}\,}
\newcommand\nsl[1] {}

\def\om            {\omega}
\newcommand\omb[1] {{\bar\omega^\star #1}}
\def\omB           {\mbox{$\bar\omega^\star$}}
\newcommand\omd[1] {{\dot\omega #1}}
\def\omD           {\dot\omega}
\def\Omeg          {\Xi}
\def\omdi          {{\omd i}}
\def\omdj          {{\omd j}}
\newcommand\omdm[1]{{\omD^{-1} #1}}
\def\omdmo         {{\omD^{-1}0}}

\def\omdnu         {{\omd\inu}}
\def\omdo          {{\omd0}}
\newcommand\ompo[1]{{\dot\omega^{#1}0}}
\newcommand\omt[1] {{\omega^\star #1}}
\def\omT           {\mbox{$\omega^\star$}}
\def\one           {\mbox{{\small $1\!\!$}{\normalsize1}}}

\def\onetor        {1,2,...\,,r}
\def\Onetor        {1{,}2{,}...{,}r}
\def\Op            {\mbox{${\rm O}^\uparrow(r{+}1{,}1)$}}
\def\otor          {0,1,...\,,r}

\def\pig           {\hat\pi^{}_{\gM}}
\def\pigl          {\pi^{}_{\gM;\el}}
\def\pil           {\hat\pi^{}_\Lambda}
\def\pilo          {\hat\pi^{}_{\omt\Lambda}}

\def\Pl            {{\cal P}^{(\lambda)}}
\def\PL            {{\cal P}_{\!\Lambda}}
\def\PLo           {{\cal P}_{\!\omt\Lambda}}
\def\q             {quantum }
\long\def\query#1{\hskip 0pt{\vadjust{\everypar={}\small\vtop to 0pt{\hbox{}%
     \vskip -13pt\rlap{\hbox to 49.0pc{\hfil{\vtop{\hsize=8pc\tolerance=6000%
     \hfuzz=.5pc\rightskip=0pt plus 3em\noindent#1}}}}\vss}}}}%
\def\rationals     {{\dl Q}}
\def\re            {0}

\def\rep           {rep\-re\-sen\-ta\-ti\-on}
\def\Rep           {Representation}
\def\resp          {respectively}

\newcommand\restrs[1]{|\raisebox{-.44em}{$\scriptstyle#1$}}
\def\rhs           {right hand side}
\def\rl            {\pi^{}_\Lambda}
\def\rlo           {\pi^{}_{\omt\Lambda}}

\def\rz            {{r+1}} 
 
\def\RZ            {{(r{+}1)}} 
\def\scs           {\scriptstyle}
\newcommand\sect[1] {\section{#1}\setcounter{equation}{0}}
\newcommand\Sect[2] {\sect{#1}\label{s.#2}
                   \ifnum\draftcontrol=1 \query{s.#2} \fi}
\def\scu           {simple current}

\def\slnh          {\mbox{$\widehat{\liefont{sl}}(N)$}}

\newcommand\smallmatrix[1] {\mbox{\footnotesize $\left(\begin{array}#1
                   \end{array}\right)$}}

\def\SOree         {\mbox{${\rm O}(r{+}1{,}1)$}}
\def\soree         {\mbox{$\liefont{so}(r{+}1{,}1)$}}

\def\sss           {\scriptscriptstyle}
\def\subseT        {\,{\subset}\,}
\newcommand\sumer[1]{{\displaystyle\sum_{#1=1}^r\,}}
\newcommand\Sumer[1]{{\sum_{#1=1}^r\,}}
\newcommand\sumne[1]{{\displaystyle\sum_{#1=1}^{N-1}}}
\newcommand\Sumne[1]{{\sum_{#1=1}^{N-1}}}

\newcommand\Sumoe[1]{{\sum_{#1=0}^{r+1}}}
\def\tauh          {\mbox{$\hat{\liefont T}_\omega$}}
\def\tauo          {\mbox{$\liefont T_\omega$}}
\def\tilD          {D_{}^{\sss[1]}}
\def\Times         {\!\times\!}
\def\ttb           {{\bar\theta}}
\def\twodim        {two-di\-men\-si\-o\-nal}
\def\U             {\mbox{\mfont U}}

\def\Uo            {\mbox{{\mfont U}$_\om$}}

\def\Um            {\U^{-1}}
\def\Umo           {\U_\om^{-1}}
 
\def\vaca          {\Omega_a}
\def\vacaL         {\Omega_\aLam}
\def\Vee           {{\nsl{\scriptscriptstyle\vee}}}
\newcommand\version[1] {\ifnum\draftcontrol=1 \typeout{}\typeout{#1}\typeout{}
                   \vskip3mm \centerline{\fbox{{\tt DRAFT -- #1 -- }
                   {\small\draftdate}}}
                   \vskip3mm \fi}

\def\vll           {v^\Lambda_\Lambda}
\def\vllo          {v^{\omt\Lambda}_{\omt\Lambda}}
\def\wrt           {with respect to }
\def\wrtt          {with respect to the }
\def\wzwt          {WZW theory}
\def\wzwts         {WZW theories}

\def\zet           {{\dl Z}}

\def\zetplus       {{\zet_{>0}}}

\def\draftdate{\number\month/\number\day/\number\year\ \ \ \hourmin }

\global\def\draftcontrol{0}
\catcode`\@=12

\documentclass[12pt]{article} \usepackage{amssymb,amsfonts}
\begin{document}
\vskip2em

\begin{flushright}  {~} \\[-15 mm] {\sf hep-th/9702195} \\[1mm]
{\sf DESY 97-023} \\[2 mm]{\sf February 1997} \end{flushright} \vskip 17mm
\begin{center} 
{\Large\bf UNIVERSAL SIMPLE CURRENT}\\[.5em]
{\Large\bf VERTEX OPERATORS}\vskip21mm
{\large J\"urgen Fuchs $^{\sss{\sf X}}$} \\[3mm]
{DESY\\[1mm] Notkestra\ss e 85\\[1mm] D -- 22603~~Hamburg}
\end{center} 
\vskip21mm
\begin{quote}{\bf Abstract}.\\
We construct a vertex operator realization for the simple current primary
fields of WZW theories which are based on simply laced affine Lie algebras \g.
This is achieved by employing an embedding of the integrable 
highest weight modules of \g\ into the Fock space for a bosonic string
compactified on the weight lattice of \g. Our vertex operators are universal 
in the sense that a single expression for the vertex operator holds 
simultaneously for all positive integral values of the level of \g. 
\end{quote}
\vfill {}\fline{} {\small $^{\sf X}$~~Heisenberg fellow} \newpage

\sect{Simple currents}

A simple current of a \twodim\
\cft\ is a primary field $\phi_J$ with \q dimension 1.
Simple currents can also be characterized by the property that their fusion 
product with any other primary field $\phi$ consists of a single primary field 
$\phi'=\phi_J\star\phi$. It follows in particular that the set of simple 
currents of a theory constitutes a subgroup of the fusion ring.
While these characterizations of simple currents in terms of the fusion rules
are both suggestive and elegant, they require to regard the primary fields as
elements of an abstract \alg ic structure, the fusion ring, and hence do not 
provide any concrete information about the action of the simple currents on 
conformal fields (via the operator product), \resp\ on the space \h\ of 
physical states. A more direct realization of simple currents in terms of the 
\rep s of the chiral \alg\ of the theory is therefore most desirable.
For general \cfts\ such a formulation is so far not available. This is 
related to the fact that the fusion product of primary fields is not 
isomorphic to the tensor product of the \rep s of the chiral \alg\ that are
carried by the primary fields,
and can be regarded as another manifestation of the difficulties which lie
behind the very concept of a conformal `field'.

In the case of WZW \cfts, the situation is different. For these theories
the chiral \alg\ is generated by an untwisted affine \kma\ \g,
so that the well-developed \rep\ theory of such \lie s allows one to analyse
the fusion rules in considerable detail. In particular, it is known that
-- with the exception of an isolated case appearing for $E_8$ level two --
the simple currents $J$ correspond to certain symmetries $\omD\equiv\omD_J$ of 
the \dyd\ of \g;
more precisely, they are in one-to-one correspondence with the maximal abelian 
subgroup of those symmetries. As a consequence, it has been possible \ci\ to
realize \scu s on the irreducible subspaces of \h, i.e.\ on the integrable
\ihwm s \hl\ of \g, as certain linear maps 
  \be  \tauo:\quad \hl \to \hlo   \labl{tauo}
associated to $\omD$, which are uniquely characterized 
(see the formul\ae\ \Erf t1 and \Erf t2 below) by their action on the \hwv\ 
and by their commutation relations with the elements of \g.

The description of WZW simple currents through these maps $\tauo$
has the advantage of clearly
displaying the underlying \rep\ theoretic structures, and it also proves to be
indispensable for various applications\,%
\futnote{Another description \cite{dolm} of WZW \scu s is in terms of twisted 
modules over a vertex operator \alg. While this does not seem to be particularly
suited for describing their action on the space of physical states, it does 
allow one to discuss extensions of the chiral \alg\ by the \scu s.}
of simple currents, e.g.\ to the resolution of field identification fixed 
points \cite{fusS4,fusS6,bant7}. In particular, by collecting the 
linear maps $\tauo$ on all of the \irmod s \hl\ which belong to the
spectrum of the theory, one arrives at a realization of the \scu s on the 
direct sum $\bigoplus_\Lambda\!\hl$ of these spaces; more specifically, 
one has to sum over all unitary \ihwm s with some fixed positive integral
value \kv\ of the level, so that the direct 
sum $\bigoplus_\Lambda\!\hl$ is isomorphic to the (chiral) state space 
$\h\equiv\h_{\gM;\el}$ of the \wzwt\ based on \g\ at level \kv.
However, in \cft\ one thinks of the irreducible subspaces $\hl\,{\subseteq}\,\h$
as arising by the action of suitable field operators on the
vacuum vector of the theory, whereas the approach of \ci\
is based entirely on the abstract description of the
\g-modules \hl\ as quotients of Verma modules. Thus the realization of WZW
\scu s through the maps \erf{tauo} is still not as field-theoretic
as one might wish.

In the present paper we obtain, for the case of simply-laced \g,
a much more concrete realization of WZW simple currents. 
To this end we embed the space \h\ into the Fock space
of a bosonic string that is completely compactified on the weight lattice of \g\
\cite{fren5,gool2,geni3}. We can then identify unitary operators \Uo\ acting on 
this big space \fock\ which, when restricted to $\h\subseT\fock$, behave 
precisely as the linear maps $\tauo$ \erf{tauo}; in other words, we construct a 
map $\tauh{:}\;\fock\,{\to}\,\fock$ such that
  \be  \tauh\restrs\hM:\quad \h \to \h\,,\ \ v\mapsto\Uo v  \Labl Uh
and
  \be  \tauh\restrs\hL = \tauo \,.  \Labl Ut
The action of \Uo\ on the vector space \fock\ induces in a natural manner an 
action on the string oscillators which operate on \fock, namely as conjugation 
$x\,{\mapsto}\,\Uo x\Umo$ of $x\iN\fock$ by \Uo. Now the string oscillators 
(supplemented by appropriate cocycle operators) can be employed to construct a
vertex operator \rep\ of \g, which by restriction to suitable subspaces of 
\fock\ provides in particular vertex
operator realizations of the integrable irreducible \hwr s $\rl$ of \g.
Moreover, the conjugation by \Uo\ supplies us with a 
realization of the WZW \scu\ as an automorphism of the \alg\ of string 
oscillators, which we can supplement in such a way that it also becomes 
an automorphism of the extension of that \alg\ by the cocycle operators.
Combining these results we arrive at the desired realization of \scu s
on the irreducible subspaces of \h.
In fact, the so obtained realization of a \scu\ constitutes an {\em inner\/} 
automorphism of the oscillator (and cocycle) \alg;\,%
\futnote{A description via inner automorphisms can also be achieved to a certain
extent within the vertex operator \alg\ formulation of \cite{dolm}.}
in contrast, when restricted to the affine \alg\ \g, \scu s act via certain 
{\em outer\/} automorphisms (see formula \Erf om). 

Besides providing a concrete action on \h,
our construction has the additional benefit of being universal
in the sense that a single expression for the vertex operator \Uo\ holds 
simultaneously for all irreducible submodules \hl\ of \h.
Accordingly, it is justified to call \Uo\ the (universal)
{\em \scu\ vertex operator\/} for the relevant \scu\ $J$.
Furthermore, this expression  for \Uo\ is in fact even valid
simultaneously for all positive integral values of the level of \g.
(This is possible because the string Fock space \fock\ contains all
the \ihwm s for all integrable \hgw s $\Lambda$ \cii.)
In other words, our realization of the \scu s not only holds for each
individual \wzwt\ based on an \aff\ \g\ at some level \kv, but it applies in
fact to the whole collection of {\em all\/} (unitary) \wzwts\ based on \g\
at all positive integral values of the level.

The rest of this paper is organized as follows.
In sections \ref{s.c} and \ref{s.l} we collect those aspects of the covariant
vertex operator construction and of the operation of the Lorentz group
on the string Fock space and string oscillators, \resp,
that are relevant to our arguments. Section \ref{s.a} is devoted to a
description of the pertinent features of the various automorphisms --
of the \dyd, of the \lie\ \g, and of its weight space \god\ -- which realize
the action of a WZW \scu. In order to relate these structures to those
discussed in the earlier sections, it is necessary to extend the action
(by conjugation) of \Uo\ on the string oscillator \alg\ to include also the
cocycle factors; this is achieved in section \ref{s.C}. In section \ref{s.t} we
put these results together, which finally allows us to demonstrate
our main result, the identification \Erf Ut of the maps $\tauo$ \erf{tauo} in 
terms of the simple current vertex operators \Uo.
We conclude with a few short remarks on possible applications and open 
questions.

\Sect{Covariant vertex operators}c

We consider a free relativistic string that is completely compactified on
a torus in such a way that the string momenta are constrained to lie on the 
weight lattice $\Lg$ of a simply laced affine \kma\ \g. Such a compactified 
string theory provides a covariant version \cite{fren5,gool2,geni3} 
of the vertex operator construction of integrable \hwr s of \g.
(In the more familiar non-covariant (Frenkel\hy Kac\hy Segal 
\cite{frka,sega,halp3}) construction one compactifies on the
root lattice of the horizontal subalgebra \gb\ of \g\ rather than on the
{\em affine\/} weight lattice.)
The state space of the theory is the Fock space \fock\ that is generated
from a collection of vacuum vectors $\vaca$ with $a\iN\Lg$
by applying the creation operators of the string oscillator \alg. 
Note that what is usually called the weight lattice of \g\ is not really
a lattice, but a one-parameter family of lattices. Namely, it consists of
the weights $\lambda\eq\tilde\lambda+c\delta$ where $\tilde\lambda$ is
an {\em integral\/} linear combination of the fundamental weights \li\ of \g\ 
and $\delta$ is the so-called null root of \g, and where $c$ is an arbitrary
{\em complex\/} number. For the covariant vertex operator construction 
one rather considers only {\em rational\/} values of $c$ \cite{geni3},
and accordingly from now on the notation $\Lg$ will refer to the
one-parameter family of lattices where the parameter is restricted to lie in
$\rationals$.

The string oscillators $\ala m\imu$ ($m\iN\zet$),
$p^\imu\,{\equiv}\,\ala0\imu$ and $q^\imu$ are the modes of the covariant
Fubini\hy Ve\-ne\-zi\-ano coordinate and momentum fields
  \be  X^\imu(z) = q^\imu-\ii p^\imu\,\ln z + \ii \sum_{m\ne0} \Frac1m\,
  \ala m\imu\,z^{-m} \qquad {\rm and} \qquad
  P^\imu(z) = \ii\,\Frac{\rm d}{{\rm d}z}\,X^\imu(z) \,.  \Labl XP
They satisfy the Heisenberg commutation relations
  \be  [q^\imu,p^\inu] = \ii\,(\Gm)^\munu\,\bfe \,, \qquad
  [\ala m\imu,\ala n\inu] = m\,(\Gm)^\munu\,\dd{m+n}0\,\bfe  \,. \Labl Hb
Here the labels $\imu,\inu$ take values in $\{0,1\Ldots\rz\}$, where $r$ is
the rank of \g, and the matrix $G$ in \Erf Hb denotes
the metric on the weight space of \g, which has Minkowskian signature.

We will denote the associative \alg\ that is generated freely by the unit
\bfe\ and by the string
oscillators $\ala m\imu$, $p^\imu$ and $q^\imu$ modulo the relations \Erf Hb
by \A. The \alg\ \A\ is a $^*$-\alg, i.e.\ is endowed with an involutive
automorphism `\,$^*$\,'. The $^*$-operation acts as $(p^\imu)^*=p^\imu$,
$(q^\imu)^*=q^\imu$, $(\ala m\imu)^*=\ala{-m}\imu$, i.e.\ in particular
it exchanges creation operators ($\ala m\imu$ with $m{<}0$) with
annihilation operators ($\ala m\imu$ with $m{>}0$).

As a basis of the weight space \god\ of \g\ we choose 
  \be  {\cal B}_0= \{\lo\} \cup \{\lib\Mid i\eq\onetor\} \cup \{\delta\}
  \,,  \Labl Bo
where \li\ and \lib\ indicate the fundamental weights of \g\ and of its
horizontal subalgebra \gb, \resp, and $\delta$ is the null root.
In the basis \Erf Bo, the metric $G$ on \god\ takes the form
  \be  G = \left( \begin{array}{ccc} 0&0&1\\[.1em] 0&\GB&0 \\[.1em] 
  1&0&0  \eear \right) ,  \labl G
i.e.\
  \be  G_\munu = \dd\imu\re\dd\inu\rz + \dd\imu\rz\dd\inu\re+\Gb\imu\inu
  \,,  \Labl G2
where $\GB$ is the (Euclidean) metric on the weight space \godb\ of \gb, i.e.\
the inverse of the \nsl{symmetrized }Cartan matrix of \gb\ (thus the allowed
range of labels of $\GB$ is $\{\onetor\}$, i.e.\ the last term
in \Erf G2 should be interpreteded as $\Gb\imu\inu\,{\equiv}\,
\Sumer{i,j}\!\delta_{i,\imu}\delta_{j,\inu}\Gb ij$).
We will write inner products \wrtt metric $G$ with a dot,
$\alpha\Cdot\beta\equiv\Sumoe\munu G_\munu\alpha^\imu\beta^\inu$
for $\alpha,\beta\iN\god$.

The covariant coordinate and momentum fields \Erf XP can be
employed to construct covariant vertex operators which,
as compared to the ordinary
vertex operator construction, have the advantage of commuting with the
Virasoro constraints and hence being manifestly physical in the sense of 
string theory. Furthermore, this construction
is applicable at arbitrary level \kv\ of the affine \lie\ \g.
Namely, for every integrable highest \g-weight $\Lambda\iN\god$, there exists
\cite{geni3} a subspace $\PL$ of the Fock space \fock\ which 
carries a vertex operator \rep\
  \be  \pil:\quad \g\to {\rm End}\,\PL  \labl{pil}
of the affine \lie\ \g.
This vertex operator \rep, which was recently also exploited in 
\cite{gekn,geni3}, will be the basis of our construction of \scu s, and 
therefore we will now describe it in some detail.

The subspace $\PL\,{\subset}\,\fock$ on which $\pil$ operates is defined as
  \be  \PL:=\!\bigoplus_{\lambda\in\Omeg(\Lambda)}\! \Pl \quad{\rm with}\quad
  \Pl:= \{v\iN\fock\,|\,
  L_0v\eq v,\; L_nv\eq0\ {\rm for\;all}\;n\,{>}\,0,\;p^\imu v\eq\lambda^\imu v\}
  \,.  \Labl PL
Here the set $\Omeg(\Lambda)$ is the weight system of the \ihwm\ 
\hl\ with \hw\ $\Lambda$, and $L_m$, $m\iN\zet$, are the 
string Virasoro operators, which are given by
  \be  L_m = \half \sum_{n\in\zet} \normord{\alpha_n\Cdot\alpha_{m-n}} \,;
  \labl{vir}
they span a Virasoro algebra of central charge $c\eq r{+}1$.\,%
\futnote{This Virasoro algebra is not related to the Virasoro algebra of the
WZW \cft\ associated to \g, which is determined through \g\ by the Sugawara
construction and has a different, level dependent, central charge.}
Moreover, every subspace $\Pl$ of $\PL$ can be obtained by applying so-called
DDF \cite{dedf} operators (which therefore constitute a spectrum generating 
\alg\ for the string) to a suitable tachyonic vacuum vector $\vaca$, namely
to one that is associated to the weight\,%
\futnote{This formula actually demonstrates that as long as one works with a 
fixed value \kv\ of the level, one needs not consider the full `lattice'
$\Lg$, where as noted above the coefficient of the null root $\delta$ can take 
arbitrary rational values $c$, but can restrict $c$ to integral multiples of 
$(2K\el)^{-1}$, where $K$ is the maximal denominator of the entries of the
matrix $\GB$ of \gb, so that one really deals with a proper lattice.\label{foo}}
  \be  a = \alam := \lambda +\el^{-1}(1-\half\lambda\Cdot\lambda)\,\delta
  \,.  \labl{alam}
Note that $\alam\Cdot\alam\eq2$ independently of $\lambda$.

The \rep\ $\pil$ 
is of level $\kv\eq\Lambda\Cdot\delta$; it is defined on a \cwb\ 
  \be  {\cal B}(\g)=\{K,D\} \cup \{H^i_m\,|\,i\eq\onetor,\;m\iN\zet\}
  \cup \{E^\er_m\,|\,\er\;{\rm a}\;\gb\mbox{-root},\;m\iN\zet\}  \labl B
of \g\ by
  \be  \bearl  \pil(K) = \delta\Cdot p \,,\qquad\
  \pil(D) = \lo\Cdot p \,,  \\{}\\[-.5em]
  \pil(H^i_m) = \izi \alv i\Cdot P(z)\,\exp[\ii m\delta\CDot X(z)] \,,
  \\{}\\[-.5em]
  \pil(E^\er_m) = \izi \normord{\exp[\ii(\er+m\delta)\CDot X(z)]}
  c_\er  \,.  \eear \Labl vo
Here $\al i$ are the simple roots of \g\ (which are orthonormal to the
fundamental weights, $\al j\Cdot\li=\delta_i^{\;j}$),
and $c_\er$ are the cocycle operators which guarantee the proper
sign factors in the commutation relations of the $\pil(E^\er_m)$.
Implementing the relation
  \be  \alb i = \al i \quad{\rm for}\ i\eq\onetor\,, \qquad
  \al0 = -\bar\theta+\delta  \ee
between the simple roots $\alb i$ ($i\iN\{\onetor\}$) of \gb\ and
$\al i$ ($i\iN\{\otor\}$) of \g\ ($\bar\theta$ denotes the highest root of \gb),
one learns in particular that the Chevalley generators of \g\ -- that is, the 
generators $H^i\,{\equiv}\,H^i_0$, $\Eipm{\equiv}\,\Eibpm$ for 
$i\iN\{\onetor\}$, and $H^0\df K\,{-}\,\Sumer j\!\av jH^j_0$,\,%
\futnote{By $\av i$ we denote the \nsl{dual }Coxeter labels of \gb, i.e.\ for 
$i\iN\{1,2\Ldots r\}$ they are the expansion coefficients of the highest 
\coro\ $\bar\theta^\Vee$ of \gb\ \wrtt simple \coro s, while $\av0\df1$.}
$\Eiopm{:=}\,\Eopm$) which are associated to the simple \g-roots -- in the 
\rep\ \erf{pil} read
  \be  \pil(H^i) = \alv i\Cdot p\,,\qquad  \pil(\Eipm)=\izi \normord{\exp[\pm\ii
  \al i\CDot X(z)]}  \quad{\rm for}\ i=\otor\,.  \ee
Moreover, one can check that these operators indeed commute with the
Virasoro \alg, $[L_m,x]\eq0$ for all $m\iN\zet$ and all $x\iN{\cal B}(\g)$.

The \g-module $\PL$ \Erf PL turns out to be highly reducible. It contains in 
particular an irreducible submodule that is isomorphic to (and will henceforth 
be identified with) the \ihwm\ \hl. 
By restriction to $\hl\,{\subset}\,\PL$, $\pil$ defines a unitary \ihwr\ 
  \be  \rl:\quad\g\to{\rm End}\,\hl  \Labl rl
of \g. The submodule \hl\ of $\PL$ can be described as the \hwm\
that is generated from a fixed vector 
  \be  \vll:=\vacaL\ \in\PL  \labl{vll}
of weight $\Lambda$
by application of (the enveloping \alg\ of) $\pil(\gm)$, with
$\gm\eq{\rm span}\{H^i_m,E^\er_m\,|\,m\,{<}\,0\}$\linebreak[0]$\,{\cup}\,
\{E^\er_0\,|\,\er\,{<}\,0\}$.
It proves to be a crucial property of the Fock space vector $\vll\iN\fock$ 
that it is in fact a vector in $\PL$, i.e.\ that it is a physical
string ground state with weight $\Lambda$ which satisfies
  \be  L_0\,\vll = \vll \,, \qquad  L_n\,\vll=0\;\ \ {\rm for}\ n>0   \Labl vL
as well as
  \be  p^\imu\,\vll = \Lambda^\imu\,\vll \quad\,\ {\rm for}\ 1\le\imu\le\rz \,.
  \Labl vp
Namely, these properties are already sufficient to verify \cii\ that $\vll$ 
both possesses the defining properties
  \be  \bearll
  \pil(K)\,\vll = \el\,\vll \,, \\[.54em]
  \pil(H^i_0)\,\vll = \Lambda^i\,\vll &{\rm for}\ 1\le i\le r \,, \\[.54em]
  \pil(\Eib)\,\vll = 0 &{\rm for}\ 1\le i\le r \,, \\[.54em]
  \pil(\Eo)\,\vll = 0   \eear \Labl hw
of a \hwv\ of \hw\ $\Lambda$ and satisfies the irreducibility
(null vector) conditions
  \be  \bearl
  (\pil(\Emib))^{\Lambda^i+1}_{}\,\vll = 0 \quad\ {\rm for}\ 1\le i\le r\,,
  \\[.54em]
  (\pil(\Emob))^{\el-\bar\Lambda\Cdot\bar\theta+1}_{}\,\vll = 0 \,.  
  \eear \Labl nw

Note that for each integrable \g-weight $\Lambda$ the space $\PL$ is by 
definition a subspace of the Fock space \fock; but in fact even the direct sum
$\bigoplus_{\Lambda\in\goD{:}\;\Lambda\cdot\delta\in\zetplus}\PL$ of these
spaces for {\em all\/} integrable weights is contained in \fock\ \cii.
Moreover, in the expressions \Erf vo for the operators $\pil(x)$ no explicit
reference to the \hw\ $\Lambda$ is necessary. Accordingly, these expressions
supply us in fact with a unitary \rep\ 
  \be  \pig = \!\bigoplus_{\scs\Lambda\in\goD \atop
  \scs\Lambda\cdot\delta\in\zetplus}\!\pil  \labl{pig}
of \g\ on the direct sum of the spaces $\PL$ for all integrable \hw s. 
By restriction, the \rep\ \erf{pig} in turn also provides a \rep\ 
  \be  \pigl= \bigoplus_{\scs\Lambda\in\goD \atop
  \scs\Lambda\cdot\delta=\el}\rl  \labl{pigl}
of \g\ on the direct sum 
$\h_{\gM;\el}\eq\bigoplus_{\Lambda;\Lambda\cdot\delta=\el}\hl$
of the corresponding \ihwm s at any fixed value $\el$ of the level, and hence
on the state space of the \wzwt\ based on \g\ at level \kv.

In short, for any fixed simply-laced \aff\ \g\ and arbitrary level $\el\iN
\zetplus$ we can regard the WZW state space $\h\,{\equiv}\,\h_{\gM;\el}$ as
embedded into the string Fock space \fock\ and $\pigl$ \erf{pigl} as a
level-\kv\ vertex operator \rep\ of \g\ on \h.

\Sect{Lorentz transformations}l

The weight space \god\ naturally carries an action of the Lorentz group
\SOree. The elements of \SOree\ are $\RZ\,{\times}\,\RZ$\,-matrices 
$\aum\equiv(\Aum\mu\nu)$ which are orthogonal \wrt the metric \erf G,
i.e.\ satisfy $\aumt G\aum\eq G$.
Furthermore, the string Fock space \fock\ carries a unitary
\rep\ of the discrete subgroup $\LO$ of the Lorentz group \SOree\ \cii\ that
leaves the weight lattice $\Lg$ of \g\ invariant. For orthochronous\,%
\futnote{In the metric used here, the subgroup $\Op$ of orthochronous
Lorentz transformations is characterized by
$\Aum00+\Aum\rz{\,\rz}-\Aum0\rz-\Aum\rz{\;\ \ 0}\,{\ge}\,2$.}
Lorentz transformations it is given by
  \be  \LOp\ni\aum \;\mapsto\; \U \equiv \U(\aum) := \daum\,
  \exp(\ii\,\gen\cdt\L) \,.  \labl U
Here 
  \be  \L^\munu:= \half \llb q^\imu p^\inu - q^\inu p^\imu - \ii \sum_{n\ne0}
  \Frac1n\, (\alpha_{-n}^\imu\alpha_n^\inu-\alpha_{-n}^\inu\alpha_n^\imu)  \lrb
  = (\L^\munu)^*  \labl L
generate infinitesimal Lorentz transformations on the Fock space \fock,
$\gen\cdt\L \equiv \sum_{\munu=0}^\rz (G\gen)_\munu\L^\munu$, and
$\gen$ is an element of the \lie\ \soree\ of \SOree\ such that
  \be  \aum = \daum\,\exp(\gen)  \Labl Mm
is a coset decomposition of $\aum$ into the product of a suitable matrix
$\daum$ with ${\rm det}\,\daum\eq{\rm det}\,\aum$ and a proper orthochronous
Lorentz transformation $\exp(\gen)$; when ${\rm det}\,\aum\eq1$, $\daum$ is
just the unit matrix, while for ${\rm det}\,\aum\eq{-}1$, $\daum\eq{-}\one$
for odd $r$ and 
\futnote{The bar in the notation $\bar\delta_{\cdot,\cdot}$ means that the 
allowed range of the indices of the Kronecker delta is from 1 to $r$, 
analogously as for the $\GB$-part of the quadratic form matrix $G$ \erf G of 
\g.}
$(\daum)_\munu\eq\db\imu\inu{+}\dd\imu\re\dd\inu\rz{+}\dd\imu\rz\dd\inu\re$
for even $r$.
(Note that $\gen\iN\soree$ satisfies $G\,\gen\,\Gm = -\gent$, or in other words,
$G\gen$ and $\gen\Gm$ are antisymmetric matrices.)

The operation \erf U of \SOree\ on \fock\ extends to an action of Lorentz 
transformations on the oscillator \alg\ \A\ by conjugation \adu,
  \be  \A\ni x \;\mapsto\; \adu(x) := \U\,x\,\Um  \,,  \labl{adu}
which by the definition of $\U$ is an inner automorphism of \A.
{}From the Heisenberg commutation relations \Erf Hb
for the string oscillator modes we deduce that 
operators with an upper index $\imu$ transform as Lorentz vectors.
For example, we have $[p^\imu,\gen\cdt\L] \eq -\ii\,(\gen p)^\imu$  
so that $p^\imu\U\eq\U(\aum p)^\imu$, and hence
  \be  \adu(p^\imu) = (\aumm p)^\imu \,, \qquad
  \adu(p_\imu) = (p\,\aum)_\imu \,. \labl{adup}
It follows e.g.\ that the Virasoro operators \erf{vir} are Lorentz singlets,
  \be  \adu(L_m) = L_m  \,.  \labl{aduL}
Also, the inner product of any operator $x^\imu\iN\A$ that is a Lorentz vector
with a weight $\lambda\iN\godb$ transforms as
  \be  \adu(\lambda\Cdot x) = \sum_\imu \lambda^\imu\,(x\aum)_\imu
  = \sum_\imu (\aum \lambda)^\imu\,x_\imu = (\aum\lambda)\Cdot x \,,  \Labl aP
i.e.\ the action on the operator can be absorbed in a transformation
  \be  \lambda^\imu \;\mapsto\; (\aum \lambda)^\imu  \ee
of the vector $\lambda$. In particular, when \aum\ leaves 
the null root $\delta\iN\godb$ invariant, i.e.\ $\aum\delta=
\delta$, then $\delta\cdot x$ is invariant under $\adu$.

\Sect{Affine \lie s}a

As already mentioned above, a
simple current $J$ of a \wzwt\ based on an affine \lie\ \g\ corresponds
(except for an isolated case occurring for $E_8$ level two) to a
symmetry of the \dyd\ of \g. Such a symmetry can be described as a permutation 
$\omD\,{\equiv}\,\omD_J$ of the index set $\{0,1,2\Ldots r\}$ which leaves
the Cartan matrix $A$ of \g\ invariant in the sense that $A^{\omdi,\omdj}
=A^{i,j}$ for all $i,j$. To such a permutation, in turn, there is associated in 
natural manner a distinguished outer automorphism $\om$ of \g. On the \cwb\ 
\erf B of \g, this automorphism $\om$ acts as \cit{eqs.\,(6.7),\,(6.8)}
  \be \bearl
  \om(K) = K \,, \\[.4em]   \om(H^i_m) = H^\omdi_m  \,, \\[.4em]
  \om(E^\er_m) = \eta_\er\,E^{\omb\er}_{m+\mer}   \eear  \Labl om
($m\iN\zet,\,i\iN\{\onetor\},\,\er\;{\rm a}\;\gb\mbox{-root}$) and
  \be  \om(D) = D - \half\Gb\omdmo\omdmo\, K + \sumer i\Gb\omdo i\,H^i_0 \,.
  \Labl oD
In \Erf om, $\mer$ denotes the integer
  \be  \mer:=\er\cdot\Lib\omdmo  \labl{mer}
(which can actually only take the values 0 or $\pm1$), while
  \be  \omb\er := - \mer\,\ttb+ \!\sum_{\scs i=1\atop \scs i\ne\omdmo}^r\!\!
  n_i\,\alb\omdi \qquad{\rm for}\;\ \er=\sumer i n_i\,\eri \,.  \labl{omb}
Moreover, $\eta_\er\iN\{\pm1\}$ are certain sign factors, which
are completely characterized by the following properties. First,
$\eta_\eri\eq1$ for all $i\eq\onetor$. And second,
whenever $\er$, $\es$ and $\er+\es$ are \gb-roots, then
  \be  \eta_\er\, \eta_\es\, \eta_{\er+\es} = \eps\er\es\,\eps{\omb\er}{\omb\es}
  \,,  \labl{eta}
where $\eps\cdot\cdot$ is the two-cocycle which constitutes the structure 
constants in the Lie bracket relations
$[E^\er,E^\es]=\eps\er\es\,E^{\er+\es}$ of \gb.\,%
\futnote{Note that $\eta_\er\eta_\es\eta_{\er+\es}\equiv\eta_{\er+\es}/
\eta_\er\eta_\es$ provides a two-coboundary, so that $\tilde\eps\er\es\df
\eps{\omb\er}{\omb\es}$ supplies an equivalent collection of structure
constants.}

Concerning the formula \Erf oD, it must be noted that the convention 
for the derivation of \g\ chosen here (and in \cii) differs from the one
used in \ci: in \ci\ the derivation was taken to be
  \be  \tilD := D - \Frac1{4N}\,\sumne m \Gb{\ompo m}{\ompo m}\,K
  + \Frac1N\,\sumer i \sumne m \Gb{\ompo m}i\, H^i_0 \,,  \labl{tilD}
where $N$ is the order of $\omD$ (and hence of $\om$). With the help of 
the identity \citq655
  \be  \sum_{m=0}^{n-1} \Gb\omdo{\omD^m i} = \half N
  \av i\,\Gb{\omdo}{\omdo}   \quad\;{\rm for}\ i\eq\onetor  \Labl55
it follows from \Erf om and \Erf oD that $\tilD$ transforms as
  \be  \om(\tilD) = \tilD + \xi_0\,K + \sumer i \xi_i\,H^i_0 \,,  \ee
where $\xi_i$ are the linear combinations
  \be  \xi_i:=\Gb\omdo i - \half\av i\,\Gb\omdmo\omdmo
  +\Frac1N\,\sumne m \llb \Gb{\ompo m}{\omdm i}-\Gb{\ompo m}i \lrb \,,
  \quad\ i\iN\{\otor\}   \Labl xi
of elements of the metric $\GB$; in particular,
$\xi_0\eq\frac1N\Sumne m \Gb{\ompo m}\omdmo-\half\Gb\omdmo\omdmo$.
By direct calculation one checks that in fact $\xi_i\eq0$ for all $i\eq\otor$
(for $i\eq0$ this was already observed in \citq656, while for $i\Ne0$
and order $N\eq2$ it is a special case of the identity \Erf55).
Thus the derivation $\tilD$ is in fact $\om$-invariant,
  \be  \om(\tilD) = \tilD  \Labl oT
(in \ci\ $\tilD$ was constructed via this property).

We will also need various properties of the map $\omB$.
First, the relation \erf{omb} can also be written as
  \be  \omb\er = \omt\er-\mer\delta \,,  \labl{omb2}
where $\omT$ is the linear map that is induced by $\om\restrs\gO$ on the weight 
space \god\ of \g\ according to
  \be  (\omt\lambda)(\om^{-1}(x))=\lambda(x)  \ee
for all $\lambda\iN\god$ and all $x\iN\go$. 
On the fundamental \g-weights, this map $\omT$ acts as \ci\ 
  \be \omT(\Lambda_{(i)}) = \Lambda_{(\omdi)} + \llb \Gb\omdmo i - \half \av i
  \, \Gb\omdmo\omdmo) \lrb\, \delta  \qquad{\rm for}\ i\eq\otor \,,  \labl{omtl}
and on the simple roots and the null root as 
  \be  \omt(\alpha^{(i)}) = \alpha^{(\omdi)} \quad{\rm for}\ i\eq\otor 
  \labl{omtal}
and
  \be  \omt(\delta) = \delta \,,  \labl{omtdelta}
\resp. Moreover, for every \gb-root $\er$, also $\omb\er$ is a \gb-root; 
in particular, from \erf{omtal} we have 
  \be  \omb(\alb i)=\left\{ \bearll - \bar\theta &{\rm for}\ i\eq\omdmo \,,
  \\[.1em] \alb\omdi &{\rm else}\,.  \eear\right.  \ee
Both $\omT$ and $\omB$ are isometries (on the weight spaces \god\ and \godb,
\resp); this implies e.g.\ that \citq639 
  \be  \Gb\omdi\omdj-\Gb ij = \av i\,\Gb\omdo\omdj+\av j\,\Gb\omdo\omdi
  -\av i\av j\,\Gb\omdo\omdo  \Labl39
for $i,j\Ne\omdmo$, and also that the vectors \erf{alam} obey
  \be  \omt\alam = \alamo  \labl{oaa}
for all \g-weights $\lambda$.

When the \g-weights are expressed in terms of the basis \Erf Bo of \god,
then the map
$\omT$ is implemented by multiplication of the components \wrt this basis
with an $\RZ\Times\RZ$\,-matrix $\aum\,{\equiv}\,\aum_\om$; the entries
of $\aum$ read
  \be  \bearll  \Aum\imu\inu =\!& \db\omdnu\imu + (\dd\inu\re-\av\inu)\,
  \dd\imu\omdo + (\Gb\omdmo\inu-\half\Gb\omdmo\omdmo\,\dd\inu\re)\,\dd\imu\rz
  \\[.5em]&
  +\, \dd\imu\re\,\dd\inu\re + \dd\imu\rz\,\dd\inu\rz \,.  \eear \labl M
Here for $\inu\iN\{\otor\}$ the numbers $\av\inu$ are the \nsl{dual }Coxeter 
labels of \g\ and $\av\rz\df0$. 

As an illustration, we display the matrix \aum\ for the case of the `basic' 
order-$N$ automorphism of $\g\eq\slnh$, for which the permutation $\omD$
acts as $i\,{\mapsto}\,i{+}1\,{\rm mod}\,N$, for the values $N\eq2$ and $N\eq3$:
  \be  \aum_{(N=2)} =\smallmatrix{{rrr} 1&0&0\\1&-1&0\\\!\!-1/4&1/2&1 } \,,
  \qquad\ \aum_{(N=3)}
        =\smallmatrix{{rrrr}1&0&0&0\\0&0&1&0\\1&-1&-1&0\\\!\!-1/3&2/3&1/3&1}
  \,. \ee
We also note that for all $N$, the determinant of these matrices is
  \be  {\rm det}\,\aum = (-1)^{N+1} \,, \ee
and that
  \be  \aum_{(N=2)} = -\exp(\gen_{(N=2)})  \quad{\rm with}\quad
  \gen = \ii\pi \cdot \smallmatrix{{rrr} 1&0&0\\1/2&0&0\\0&-1/4&-1 } .  \ee

The following general properties of the matrix \aum\ -- valid for arbitrary 
untwisted affine \lie\ \g\ (i.e.\ even for non-simply laced ones) and for
arbitrary simple current symmetries $\omD$ of the \dyd\ of \g\ --
can be directly deduced from its
definition. First, $\aum\delta\eq\delta$, in accordance with \erf{omtdelta}.
Second, \aum\ is unipotent; its order equals the order $N$ of $\omD$:
  \be  \aum^N=\one \,. \ee
And third, $\aum\iN\LOp$; more precisely, $\aum$ is a Lorentz transformation, 
it is orthochronous, and and it maps the weight lattice $\Lg$ to itself.

That $\aum$ is orthochronous follows from
$\Aum00+\Aum\rz{\,\rz}-\Aum0\rz-\Aum\rz{\;\ \ 0}\eq2\,{+}\,\Gb\omdmo\omdmo/2$.
To deduce that $\aum \iN\SOree$, i.e.\ that it satisfies $\aumt G\aum\eq G$ 
with the metric $G$ given by \erf G,
one proceeds as follows. With the help of the identity
\Erf39 one derives from the explicit expressions \erf M and \Erf G2 that
  \be  \bearll  (\aumt G\aum - G)_{\imu,\inu} = \!&\;
  \dd\imu\re\, (\Gb\omdmo\inu+\Gb\omdo{\omd\inu}-\av\inu\,\Gb\omdo\omdo)
  \\[.6em]& \!\!\!\!+\,
  \dd\inu\re\, (\Gb\omdmo\imu+\Gb\omdo{\omd\imu}-\av\imu\,\Gb\omdo\omdo)
  \\[.6em]& \!\!\!\!+\,
  \dd\imu\re\, \dd\inu\re\, (\Gb\omdo\omdo-\Gb\omdmo\omdmo)  \,. \eear \ee
For order $N\eq2$, the last term on the \rhs\ is manifestly zero, while the
other terms vanish owing to the identity \Erf55.
For larger order, it is again easily seen that the last term on the \rhs\ 
vanishes, namely as a consequence of the fact that 
in these cases the fundamental weights $\Lib\omdo$ and
$\Lib{\omdmo}$ are each others conjugates; that the other terms vanish 
as well can be checked case by case.

Finally, to see that $\aum$ maps $\Lg$ to itself one must first note that
the relation between \g-weights $\lambda$ and the associated vectors
$\alam$ \erf{alam} is not one-to-one, but rather $\alamc\eq\alam$ for all
$c\iN\complex$\,. Moreover, the isomorphism class of the \g-module \hl\ also
depends on the highest weight $\Lambda\iN\Lg$ only modulo $\complex\delta$.
Taken together, this implies that in the construction above we can prescribe
the $\delta$-components of the weights $\Lambda$ in the construction of
section \ref{s.c} as we like. In particular we can choose $\Lambda$ in such 
a way that its coefficient in $\delta$-direction is an integral multiple of
$(2K\el)^{-1}$. Doing so, comparison with the formula \erf{omtl} (see also
footnote \ref{foo}) shows that together with $\Lambda$, also $\aum\Lambda$
lies on the lattice $\Lg$.

\Sect{Cocycle factors}C

We would like to describe the action of the conjugation $\adu$ \erf{adu}
for Lorentz transformations which leave the null root $\delta$ invariant
not only on the string oscillators, but also on the vertex operators \Erf vo.
It follows immediately from the result \Erf aP that on the generators
$K$, $D$ and $H^i_m$ of the vertex operator \rep\ $\pil$ this action reads
  \be \bearl  \adu\circ\pil(K) = \pil(K)  \,, \\[.4em]
  \adu\circ\pil(D) = (\aum\lo) \Cdot p  \,, \\[.4em]
  \adu\circ\pil(H^i_m) = \izi (\aum\alv i)\Cdot P(z)\,\exp[\ii m\delta
  \CDot X(z)] \,. \eear  \labl{adupi} 
On the other hand, the action on the generators $E^\er_m$ requires more work.
Using the identity $c_\er^2\eq\bfe$ that is satisfied by the cocycle operators
$c_\er$, we can write
  \be \adu\circ\pil(E^\er_m) 
  = \pil(E^{\auM\er}_m)\Cdot c_{\auM\er}^{}\,\adu(c_\er) \,.  \Labl0e
Thus in order to determine the action on the generators $E^\er_m$ we need to
know the action of $\adu$ on the cocycle operators. This action, however,
is not determined yet, because the cocycle operators constitute an
additional input which is not provided by the string oscillators.

In other words, it is necessary to extend the definition of $\adu$ from the
algebra \A\ to its extension by the cocycle operators. We will not attempt to
achieve this in full generality, but be content to define the action in the
case of the specific Lorentz transformations \erf M, i.e.\ for $\adu\eq\ado$.\,%
\futnote{Another case which can be treated is the one where the
Lorentz transformation is an element of the translation subgroup of
the Weyl group of \g. In that case it is consistent to require that the
cocycle operators are $\adu$-invariant \cii.}
To this end we 
recall (see e.g.\ appendix B of \cite{dogm2}) that the cocycle operators
for all \gb-roots (and more generally, for every element of the root lattice of
\gb) can be described in terms of certain gamma matrices $\c i\,{\equiv}\,
\c\eri$ associated to
the simple roots of \gb. We will define an action of $\ado$ on these gamma
matrices, which then immediately supplies the action on the cocycle operators.

The gamma matrices satisfy the relations
  \be  \c i\c j=(-1)^{A^{i,j}}_{}\c j\c i \qquad{\rm for}\;\ i,j\eq\onetor
  \labl{cicj}
(here $A$ is the Cartan matrix of \g, and we have used that $\erI\Cdot\erJ\eq
A^{i,j}$ for $i,j\eq\onetor$). The extension to arbitrary roots (\resp\
elements of the root lattice of \gb) $\er$ is based on the prescription
  \be  \c\er := \ETA_\er\, (\c1)^{n_1}_{}(\c2)^{n_2}_{}\cdots(\c r)^{n_r}_{}
  \qquad {\rm when}\;\ \er=\sumer i n_i\,\eri \,,  \Labl cc
which implies $\c\er\c\es\eq(-1)^{\er\cdot\es}\c\es\c\er$.
Here $\ETA_\er$ is a sign factor which for any $\er$ can be prescribed 
arbitrarily; in particular, it can be chosen in such a way that
  \be  \c\er\c\es = \eps\er\es\,\c{\er+\es}  \labl{cce}
for all $\er,\es$, where $\eps\cdot\cdot$ is the two-cocycle appearing in
formula \erf{eta}, and we will stick to this choice from now on.

We now define $\ado$ on the associative \alg\ \AC\ that is freely generated by 
the gamma matrices $\c i$ ($i\eq\onetor$) modulo the relations \erf{cicj} as 
follows. We set
  \be  \ado(\c i) := \c\omdi \equiv \c{\omb\eri} \qquad{\rm for}\;\ i\eq\otor 
  \,,  \labl{aduci}
where we introduced the convention that
  \be  \c i = \left\{ \bearll \c\eri & {\rm for}\ i\eq\onetor \,, \\[.1em]
  \c{-\ttb} & {\rm for}\ i\eq0  \,, \eear\right.  \Labl ci
and we impose the homomorphism property
  \be  \ado(\c\er\c\es)=\ado(\c\er)\,\ado(\c\es)  \,.  \labl{crcs}

In order for the definition 
\erf{aduci} to be consistent and to consistently extend to the whole \alg\ \AC,
it is necessary and sufficient that, first, for $i,j\eq\onetor$ \erf{aduci} 
is compatible with the relations \erf{cicj} of \AC, and second, that
the definition of $\ado(\c0)$ is compatible with \Erf cc.
These requirements are indeed satisfied: The first follows as
  \be  \ado(\c i\c j)=\c\omdi\c\omdj=(-1)^{A^{\omdi,\omdj}}_{}\c\omdj\c\omdi
  =(-1)^{A^{i,j}}_{}\ado(\c j\c i)  \ee
by the fundamental property $A^{\omdi,\omdj}\eq A^{i,j}$ of the permutation
$\omD$. The second property is a direct consequence of the fact that the
relation \erf{cicj} continues to hold for $i\eq0$, which in turn follows from
$-\sum_{i=1}^r\av iA^{i,j}\eq A^{0,j}$ 
(recall that $\ttb\eq\sum_{i=1}^r\av i\erI$).

The extension of the prescription \erf{aduci} to the whole \alg\ \AC\ is
achieved by implementing the property \erf{crcs}. We find that this leads to 
the formula 
  \be  \ado(\c\er)=\eta_\er^{}\,\c{\omb\er}^{}   \labl{adu.c}
for all \gb-roots $\er$, where $\eta_\er$ is the sign factor that was 
introduced before the relation \erf{eta}. (This is most conveniently proven 
by induction on the
height of $\er$, i.e.\ on the number $\Sumer i n_i$ of simple roots
`contained' in $\er\eq\Sumer i n_i\,\eri$. Namely, for $\er$ of height larger
than one, one can write $\er\eq\es+\erI$ for some $i\Ne\omdmo$. Then the height
of $\es$ is smaller than the one of $\er$, so that by the induction assumption
we have $\,\ado(\c\es)\eq\eta_\es\c{\omb\es}$. Hence
  \be  \ado(\c\er)=\eps\es\eri\,\ado(\c\es\c i)
  =\eps\es\eri\eta_\es^{}\,\c{\omb\es}\c\omdi
  =\eps\es\eri\eta_\es^{}\,\eps{\omb\es}{\omb\eri}\c{\omb(\es+\eri)}  \,, \ee
which by $\eta_\erI\eq1$ and by the identity \erf{eta} satisfied by the signs
$\eta_\er$ reduces to the equality \erf{adu.c}.)

When the cocycle operators are defined via the gamma matrices as in
\cite{dogm2}, then the result \erf{adu.c} translates into an analogous
formula for the cocycle operators themselves:
  \be  \ado(c_\er)=\eta_\er^{}\,c_{\omb\er}^{} \,.  \labl{aduc}
It is straightforward to check that this prescription is indeed compatible
with all properties of the cocycle operators. For instance, their basic
property reads
  \be  \eE^{\er\cdot q}c_\er\,\eE^{\es\cdot q}c_\es = \eps\er\es\,\eE^{(\er+\es)
  \cdot q}\,c_{\er+\es} \,;  \ee
upon application of $\ado$, the \lhs\ becomes
  \be  \eta_\er\eta_\es\,\eE^{\omb\er\cdot q}c_{\omb\er}\,\eE^{\omb\es\cdot q}
  c_{\omb\es} = \eta_\er\eta_\es\,\eps{\omb\er}{\omb\es}\,\eE^{\omb(\er+\es)
  \cdot q} c_{\omb\er+\omb\es} \,,  \ee
and by \erf{eta} this coincides with what is obtained
by acting with $\ado$ on the \rhs.

We can now come back to our original problem to compute $\ado\Circ\pil
(E^\er_m)$. Combining the formul\ae\ \Erf0e and \erf{aduc} we obtain
(recall that $\aum\er\eq\omb\er$)
  \be \ado\circ\pil(E^\er_m) = \eta_\er^{}\,\pil(E^{\auM\er}_m) \,.  \labl e

\Sect{The \scu\ vertex operators}t

Putting the results of the previous sections together, we are finally in a 
position to obtain the interpretation of the unitary operators \Uo\ as
vertex operators for simple currents.
As already mentioned in the introduction, in \wzwts\ one can realize \scu s
by the linear maps $\tauo$ \erf{tauo} between the \ihwm s \hl\ and \hlo\ of \g. 
As the notation $\tauo$ indicates, this \scu\ map corresponds to
the outer automorphism $\om$ \Erf om of \g. In fact, the map $\tauo$ 
is completely characterized by its action on the \hwv\ of \hl, which reads
  \be  \tauo:\quad \vll \mapsto \vllo \,,  \Labl t1
and by the $\omega$-{\em twining property\/} which says that
  \be  \tauo \circ \rl(x) = \rlo(\omega(x)) \circ \tauo  \Labl t2
for all $x\iN\g$, where $\rl$ denotes the \ihwr\ \Erf rl of \g.
Because of \Erf t2, $\tauo$ was called an $\omega$-{\em twining map\/} in \ci.
  
\vskip.3em\noindent
We will now demonstrate: \begin{quote} The $\omega$-twining map $\tauo$ 
\erf{tauo} between \ihwm s \hl\ and \hlo\ of \g\ 
can be realized as the unitary vertex operator $\Uo\,{\equiv}\,\U(\aum)$ of the
Lorentz transformation $\aum\,{\equiv}\,\aum_\om$ \erf M on the string Fock 
space \fock. \\More precisely, by restriction of the linear map
  \be  \bearll \tauh:\ & \fock \to \fock \\[.3em]& \,v\, \mapsto \Uo\,v 
  \eear \Labl tf
on \fock\ to the subspace $\hl\,{\subseteq}\,\PL\,{\subseteq}\,\fock$ one 
reproduces the map $\tauo$. \end{quote}

\noindent
To prove this statement, we show that the map \Erf tf satisfies the analogues
  \be  \tauh(\vll)=\vllo  \Labl h1
and
  \be  \tauh \circ \pil(x) = \pilo(\om(x)) \circ \tauh  \Labl h2
of \Erf t1 and \Erf t2, and that these properties are
preserved by the restriction to \irmod s.

To verify \Erf h2, we first observe that the formul\ae\ \erf{adupi} and \erf e 
hold for $\aum\,{\equiv}\,\aum_\om$, and that we can rewrite those equations as
  \bea
  && \ado\circ\pil(D) = (\omt\lo) \cdot p       \,, \label{adupi1}\\[.4em]
  && \ado\circ\pil(K) = \pilo(K) \,, \label{adupi2}\\[.2em]
  && \ado\circ\pil(H^i_m) = \izi (\omt{\alv i})\cdot P(z)\,\exp[\ii m\delta
     \CDot X(z)] = \pilo(H^\omdi_m)\,, \label{adupi3}\\[.2em]
  && \ado\circ\pil(E^\er_m) = \eta^{}_\er\, \pilo(E^{\omb\er}_{m+\mer}) \,.
     \label{adupi4} 
  \labla{adupi1..4}
Here we recalled that in the vertex operator \rep\ \Erf vo no explicit
reference to the \hw\ $\Lambda$ is made and that $\auM\Lambda\eq\omt\Lambda$.
In addition, in order to arrive at \erf{adupi3} we implemented the identity 
\erf{omtal} for $\omt{\alv i}$, and for \erf{adupi4} we used the fact
that according to the definitions \Erf vo of $E^\er_m$ and \erf{mer}
of the integer $\mer$ we have to identify
  \be  E^{\auM\er}_m \equiv E^{\omt\er}_m =E^{\omb\er+\mer\delta}_m
  = E^{\omb\er}_{m+\mer} \,.  \ee
Also note that in the first place the \rhs\ of \erf{adupi3} strictly makes 
sense only for $i\Ne\omdmo$; however, it extends to the case $i\eq\omdmo$
(and, likewise, to $i\eq0$) when one adopts the convention to write
  \be  H^0_m := \delta_{m,0}^{}\,K-\sumer j \av j\,H^j_m \,.  \ee

The results \erf{adupi3} and \erf{adupi4} tell us
in particular that on the Chevalley generators of \g\ we simply have
  \be  \adu\Circ\pil(H^i)=\pilo(H^\omdi)\,, \quad
  \adu\Circ\pil(\Eipm)=\pilo(\Eomdipm)\quad  {\rm for\ all}\ i=\otor\,. \ee 
Furthermore, by noticing that $\av\omdo\eq1$, so that $\omt\lo-\lo \eq \loo
-\av\omdo\lo \eq \loob \eq \Sumer i\! \Gb\omdo i\alb i$,
and using the identity $\xi_i\eq0$ for the numbers $\xi_i$ defined in
\Erf xi, one deduces from the results \erf{adupi1} to \erf{adupi3}
that the linear combination $\tilD$ \erf{tilD} of generators of \g\ obeys
  \be  \ado\circ\pil(\tilD) = \lo\Cdot p = \pilo(\tilD) \,. \labl{adupi5}

Now comparing the formul\ae\ \erf{adupi2} to \erf{adupi4} and \erf{adupi5}
with the action \Erf om and \Erf oT of the automorphism $\om$ of \g, we 
conclude that the equality
  \be  \ado\circ\pil(x) = \pilo(\om(x))   \ee
holds for all $x\iN\{K,\tilD\} \,{\cup}\, \{H^i_m|i\eq\Onetor,\,m{\in}\zet\}
\,{\cup}\,\{E^\er_m|\er\,{\rm a}\,\gb\mbox{-root},\,m{\in}\zet\}$, and hence
by linearity of $\om$ for all $x\iN\g$. Rewriting this identity as
  \be  \Uo\,\pil(x) = \llb\ado\circ\pil(x)\lrb\,\Uo = \pilo(\om(x))\,\Uo \,,
  \Labl Hx
we learn that the map $\tauh$ \Erf tf indeed satisfies the analogue
\Erf h2 of \Erf t2.

Next we recall that the vector $\vll\iN\hl$ lies in the subspace $\PL$ of
\fock\ and is characterized by the properties \Erf vL and \Erf vp.
Combining the first of those properties with \erf{aduL} we obtain
  \be  L_0\,\Uo\vll = \Uo\vll \qquad{\rm and}\qquad L_n\,\Uo\vll=0\;\ \
  {\rm for}\ n>0 \,,  \ee
while making use of \erf{adup} it follows that
  \be  p^\imu\,\Uo\vll = \Uo\,(\aum p)^\imu\,\vll
  = (\aum\Lambda)^\imu\,\Uo\vll = (\omt\Lambda)^\imu\,\Uo\vll  \,. \ee
Thus we can conclude that the vector $\Uo\vll\iN\fock$ lies in fact in $\PLo$,
i.e.\ is a physical ground state with \g-weight $\omt\Lambda$; in particular,
taking also into account that the associated vectors $\alam$ satisfy \erf{oaa},
$\Uo\vll$ can be regarded as the \hwv\ $\vllo$ of the irreducible submodule
$\hlo\,{\subset}\,\PLo$. Hence \Erf h1 is satisfied, as claimed.
(It is straightforward to check directly that the \hwv\ properties hold.
Namely, it follows immediately from \Erf Hx that
  \be  \pilo(x)\,\Uo\vll = \Uo\,\pil(\om^{-1}(x))\,\vll  \ee
for all $x\iN\g$. In particular, we have
  \be  \bearll
  \pilo(H^i)\,\Uo\vll = \Uo\,\pil(H^{\omdm i}_0)\,\vll =
    \Lambda^{\omdm i}\,\Uo\vll \equiv (\omt\Lambda)^i_{}\,\Uo\vll
    & {\rm for}\ 0\le i\le r \,, \\[.5em]
  \pilo(\Ei)\,\Uo\vll = \Uo\,\pil(\Eomdmi)\,\vll = 0
    & {\rm for}\ 0\le i\le r \,, \eear \ee
which shows that $\Uo\vll$ possesses the defining properties of a \hwv\ of
weight $\omt\Lambda$. Similarly, by
  \be  \llb\pilo(\Emi)\lrb_{}^{(\omt\Lambda)^i+1}\Uo\vll = \Uo\,\llb 
  \pil(\Emomdmi)
  \lrb_{}^{\Lambda^{\omdm i}+1}\vll = 0 \quad\ {\rm for\ all}\;\ i\eq\otor  \ee
one verifies that the irreducibility (null vector) conditions are satisfied.)

Finally,
it is clear that the properties \Erf h1 and \Erf h2 survive the restriction to 
the irreducible submodules \hl\ of $\PL$. More precisely, because of 
$\vll\iN\hl$, \Erf t1 is immediately implied by \Erf h1, while the results 
\erf{adupi2}\,--\,\erf{adupi4} and \erf{adupi5} show that
$(\tauh\Circ\pil(\cdot))\restrs\hLo \eq \tauo\Circ\rl(\cdot)$ and
$(\pilo(\cdot)\Circ\tauh)\restrs\hLo \eq \rlo(\cdot)\Circ\tauo$, so that
upon identifying $\tauo\df\tauh\restrs\hL$ as in \Erf Ut, the property
\Erf h2 reproduces \Erf t2.
This finally concludes our proof.

\sect{Outlook}

In this paper we have presented a realization of the simple currents of a 
simply-laced WZW
\cft\ as vertex operators acting on the Fock space of a compactified string
theory. Besides providing for the first time a concrete field-theoretic
description of these primary fields, this construction may also have
applications in other areas. For example, by studying the explicit form
of the bases of the \ihwm s\ of \g\ in terms of the string oscillators,
one could extract informations about the structure of the operator product
algebra. Also, by organizing the whole string Fock space \fock\ into orbits
\wrtt action of the \scu\ automorphisms $\tauh$ and 
using the connection \cite{geni,geni3} between the compactified string
theory and the root multiplicities of hyperbolic \kma s, it should be
possible to find relations between the multiplicities of roots which
correspond to different \aff\ modules.

It is also interesting to note that to the \scu\ symmetries $\omD$ of the
\dyd\ of \g\ there are not only associated the outer
automorphisms $\om$ \Erf om of \g, but in fact isomorphism classes of outer 
modulo inner automorphisms. The inner automorphisms of \g\ correspond to the 
Weyl group $W$ of \g. Now usually the abelian subgroup of $W$ that is
isomorphic to the coroot lattice is realized in a way rather different from
the form \Erf om of the outer automorphisms considered here. However, as
described in detail in \cii, these Weyl transformations can also be described 
as elements of the subgroup $\LOp$ of the Lorentz group 
$\SOree$, and hence on precisely the same footing as the
outer automorphisms. This should prove useful in applications to coset
\cfts, where in order to study the field identification procedure, it is
necessary (except for generalized diagonal cosets \cite{fusS4}) to deal also
with inner automorphisms of the relevant affine subalgebra of \g.

Finally we should mention that in our discussions we have followed the 
habit of talking about `the' \hwv\ of a \hw\ \g-module \hl. Strictly
speaking, the \hwv\ is unique only up a to scalar multiple.
For our construction this is quite irrelevant as long as the weight
$\omT\Lambda$ is different from $\Lambda$. In contrast, when $\Lambda$ is a
fixed point of the \scu, i.e.\ when $\omT\Lambda\eq\Lambda$ (which for
integrable weights is only possible for a proper subset of the allowed levels
\kv\ of \g), then $\tauh$ is an {\em endo\/}morphism of $\PL$ and
there is room for an arbitrary phase. To get a complete picture also
for such modules will therefore require further study.

\vskip3em\small
\noindent
{\small{\bf Acknowledgements.}\\[.3em] It is a pleasure to thank 
H.\ Nicolai, R.W.\ Gebert and C.\ Schweigert for helpful discussions 
and for valuable comments on the manuscript.
\vskip3em

 \newcommand\Bi[1]    {\bibitem{#1}}
 \newcommand\Erra[3]  {\,[{\em ibid.}\ {#1} ({#2}) {#3}, {\em Erratum}]}
 \newcommand\BOOK[4]  {{\em #1\/} ({#2}, {#3} {#4})}
 \newcommand\vypf[5]  {{#1} [FS{#2}] ({#3}) {#4}}
 \newcommand\J[5]   {{\sl #5}, {#1} {#2} ({#3}) {#4} }
 \newcommand\Prep[2]  {{\sl #2}, preprint {#1}}
 \def\wb{\,\linebreak[0]} \def\wB{$\,$\wb}
 \def\jf    {J.\ Fuchs}
 \newcommand\inBO[7]{{\sl #7}, in:\ {\em #1}, {#2}\ ({#3}, {#4} {#5}), p.\ {#6}}
 \newcommand\voms[2] {\inBO{Vertex Operators in Mathematics and
            Physics {\rm [M.S.R.I.\ publication No.\ 3]}} {J.\ Lepowsky,
            S.\ Mandelstam, and I.M.\ Singer, eds.} \SV\NY{1985} {{#1}}{{#2}}}
 \def\anop  {Ann.\wb Phys.}
 \def\npbF  {Nucl.\wb Phys.\ B\vypf}
 \def\nupb  {Nucl.\wb Phys.\ B}
 \def\phlb  {Phys.\wb Lett.\ B}
 \def\comp  {Com\-mun.\wb Math.\wb Phys.}
 \def\inma  {Invent.\wb math.}
 \def\lemp  {Lett.\wb Math.\wb Phys.}
 \def\phrd  {Phys.\wb Rev.\ D}
 \def\A       {Algebra}
 \def\alg     {algebra}
 \def\Be     {{Berlin}}
 \def\BIR    {{Birk\-h\"au\-ser}}
 \def\Ca     {{Cambridge}}
 \def\con     {conformal\ }
 \def\CUP    {{Cambridge University Press}}
 \def\furu    {fusion rule}
 \def\fts     {field theories}
 \def\ide     {identification}
 \def\Infdim  {Infinite-dimensional}
 \def\NY     {{New York}}
 \def\nn      {$N=2$ }
 \def\oa      {operator algebra}
 \def\Q       {Quantum\ }
 \def\q       {quantum\ }
 \def\qg      {quantum group}
 \def\Rep     {Representation}
 \def\SV     {{Sprin\-ger Verlag}}
 \def\syms    {sym\-me\-tries}
 \def\va      {Vira\-soro algebra}
 \def\wzw     {WZW\ }

\end{document}